\documentclass[prd,onecolumn]{revtex4}
\usepackage{amsmath,amssymb,graphics,epsfig,subfigure}
\usepackage{color}
\usepackage[colorlinks,linkcolor=red,anchorcolor=red,citecolor=green]{hyperref}
\usepackage{setspace}
\usepackage{booktabs}
\usepackage{mathtools}
\usepackage{float}

\begin{document}
	\thispagestyle{empty}
\begin{center}
	\title{Thermodynamic supercriticality and complex phase diagram for the AdS black hole}
	
	\author{Zhen-Ming Xu$^{1,2,3,}$\footnote{E-mail: zmxu@nwu.edu.cn} and Robert B. Mann$^{4,}$\footnote{E-mail: rbmann@uwaterloo.ca}
		\vspace{6pt}\\}
	
	\affiliation{$^{1}$School of Physics, Northwest University, Xi'an 710127, China\\
		$^{2}$Shaanxi Key Laboratory for Theoretical Physics Frontiers, Xi'an 710127, China\\
		$^{3}$Peng Huanwu Center for Fundamental Theory, Xi'an 710127, China\\
        $^{4}$Department of Physics $\&$ Astronomy, University of Waterloo, Waterloo, Ont. Canada N2L 3G1}
	
	\begin{abstract}
In this study, we extend the application of the Lee-Yang phase transition theorem to the realm of AdS black hole thermodynamics, thereby deriving a comprehensive complex phase diagram for such systems. Our research augments extant studies on black hole thermodynamic phase diagrams, particularly in the regime above the critical point, by delineating the Widom line of AdS black holes. This boundary segregates the supercritical domain of the phase diagram into two disparate zones. As the system traverses the thermodynamic crossover within the supercritical region, it undergoes a transition from one supercritical phase to another, while maintaining the continuity of its thermodynamic state functions.
This behavior is fundamentally different from that below the critical point, where crossing the coexistence line results in discontinuities of thermodynamic state functions. The Widom line enables a thermodynamic crossover between single-phase states without traversing the spinodal that emerges in the critical region.
	\end{abstract}

\maketitle
\end{center}

\section{Motivation}

Some our best clues as to how gravity and quantum theory can be reconciled come from the physics of black holes. In particular, their
thermodynamics~\cite{Bekenstein1973,Hawking1976,Bardeen1973} offers revealing perspectives into this notorious problem.  In recent years black hole thermodynamics has  experienced remarkable growth, greatly deepening our comprehension of the physics governing strong gravitational fields~\cite{Kubiznak2017,Visser2022,Wei2022a,Xu2023}.  Even more recently the significant accumulation of gravitational wave observations and the rapid advancement of quantum computing have enabled scientists to acquire further knowledge concerning the laws of black hole mechanics and their dynamics within curved spacetime via experimental simulations, thereby providing indirect reinforcement to  black hole thermodynamics~\cite{Isi2021,Shi2023,Wang2024}. Research on black hole thermodynamics has primarily focused on
phenomena up to the critical point, analyzing the rich landscape of phase structures below the critical point and the scaling behaviour in its vicinity.  Little attention has been to exploring which states exist above the critical point.

In conventional thermodynamic systems recent research has increasingly explored supercritical behaviour in systems such as van der Waals fluids and water, with significant advancements both theoretically and experimentally~\cite{Xu2005,Simeoni2010,Ouyang2024}. Beyond the critical point, liquids and gases become indistinguishable, forming a single fluid phase. However, certain thermo-physical properties, such as specific heat at constant pressure $C_P$ and the isothermal compression coefficient $\kappa_T$, exhibit maxima along a line extending from the critical point. This line is generally considered to be the Widom line, which  defines the locus of maximum correlation length that extends into the single fluid phase beyond the critical point~\cite{Xu2005,Widom1972}. In practice, the Widom line itself is difficult to find, and therefore response function maxima are regularly used to estimate its location. In~\cite{Simeoni2010}, Simeoni and colleagues used inelastic X-ray scattering measurements and molecular dynamics simulations to study supercritical fluid argon. They found that, despite the absence of a sharp phase transition, supercritical fluids display distinct liquid-like or gas-like behavior depending on a function of density. These differences arise from nanoscale density fluctuations generated by complex interactions between particles in the fluid. Crossing the Widom line triggers a sharp transition, effectively dividing the supercritical region into two regimes---gas-like and liquid-like---echoing behavior seen below the critical point, although not connected by a first-order phase transtion. The study's findings extend to other systems, including neon, oxygen, nitrogen, molten alkali metals, and water, offering broader insights into fluid dynamics~\cite{Xu2005,Simeoni2010}.

The studies on real supercritical fluids have demonstrated the existence of distinct phases with significantly different physical properties within the supercritical region. Consequently, identifying the boundaries between these phases has become an important research focus. However, current definitions of these boundaries remain ambiguous. Several approaches~\cite{Xu2005,Brazhkin,Prescher,Cockrell,Vega,Nishikawa,Matsugami} have been proposed to characterize them, including the Widom line (the locus of maximum correlation length), the Frenkel line (the particle motion transitions from oscillatory to diffusive), the Fisher-Widom line (the onset of oscillatory decay in the radial distribution function), the Nishikawa line (the ridge of density fluctuations on the $P-T$ phase diagram) etc. Among these, the Widom line is the most widely accepted descriptor.

Since the analysis of black hole thermodynamic phase transitions is inspired by  methodologies used in conventional thermodynamic systems,  here we carry out an investigation of  supercritical phenomena in black hole thermodynamics, aiming to theoretically uncover more information about black hole thermodynamic systems in the supercritical region. The rationale for considering supercritical phenomena in black hole thermodynamics is that below the critical point, when a system crosses the coexistence line transitioning from one single phase to another (such as in liquid-gas phase transitions or the small-large black hole phase transitions), it can encounter the spinodal line, which demarcates metastable states of supercooling or superheating.  In this sense the coexistence line is not an ideal thermodynamic crossover for single-phase transitions.  Methodologically we shall incorporate supercritical behaviour into black hole thermodynamics by employing   Lee-Yang phase transition theory.   Once the framework for analyzing supercritical behaviour  is established, the physical understanding of the associated supercritical characteristics will logically follow.

\section{Supercriticality for the AdS black hole}
The Lee-Yang phase transition theory posits that the behaviour of zeros in the grand partition function, known as Lee-Yang zeros, governs the occurrence of phase transitions. Specifically, non-analytic changes in the system's state functions arise exclusively when complex Lee-Yang zeros converge onto the real axis within the thermodynamic limit~\cite{Yang1952,Lee1952}. Building upon this foundation, Fisher extended the Lee-Yang phase transition theory to the canonical ensemble, introducing the concept of Fisher zeros for the complex temperature~\cite{Fisher1965}.

In the context of black hole thermodynamics, a crucial insight emerges from the relationship between the Euclidean path integral and the black hole partition function, which plays a pivotal role in understanding their thermodynamic properties~\cite{Gibbons1977}. In the Euclidean approach to quantum gravity, we can relate to the partition function $Z$ through the bridge equation $G=TI=-T\ln Z$, where $G$ is the Gibbs free energy, $T$ is the temperature of the canonical ensemble, and $I$ is Euclidean action. Hence the Lee-Yang phase transition theory for the black hole thermodynamics is equivalent to the following statement,
\begin{equation}\label{rule}
\text{Zeros of Partition function}\quad \xLeftrightarrow{\quad\text{$G=-T\ln Z$}\quad} \quad \text{Singularities of Gibbs free energy}.
\end{equation}
Here we demonstrate that, since the temperature of a black hole thermodynamic system remains non-singular, except for extreme black holes, which correspond to zero temperature, the zeros of the partition function are exclusively associated with the singularities of the Gibbs free energy.

\begin{figure}[htbp]
	\centering
		\includegraphics[width=120 mm]{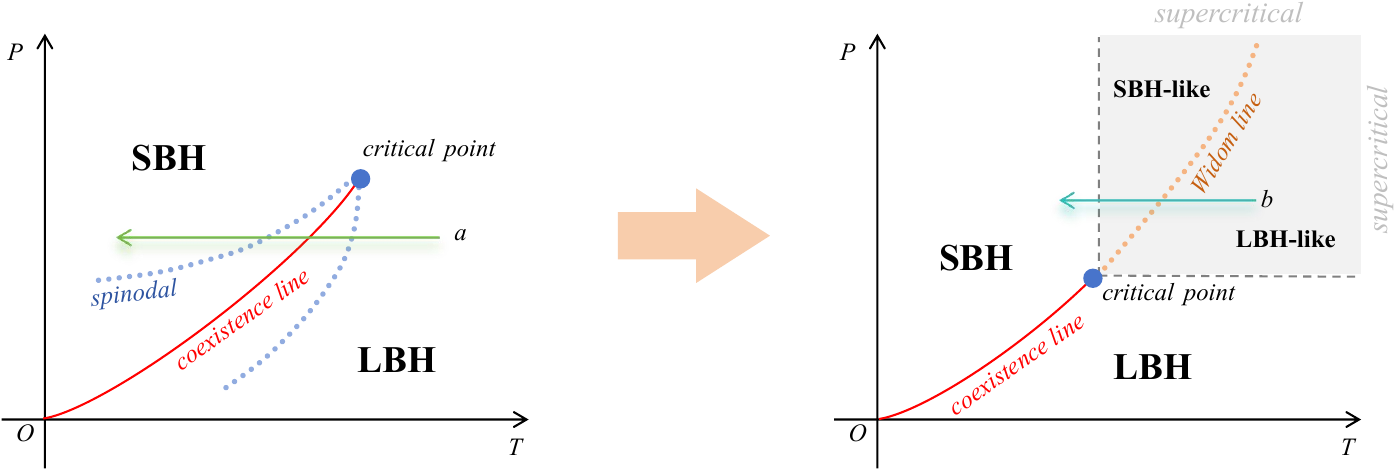}
	\caption{A schematic picture of the phase diagram of a typical  charged AdS black hole thermodynamic system. The left panel shows the phase diagram of a large black hole (LBH) and a small black hole (SBH) undergoing a phase transition (similar to a gas-liquid phase transition). The right panel shows the supercritical phenomenon of the charged  AdS black hole, analogous to conventional supercritical fluids.}
	\label{fig1}	
\end{figure}

For the charged AdS black hole thermodynamic system, the phase diagram (Fig.~\ref{fig1})  exhibits a striking analogy to the phase transition behavior of van der Waals fluids~\cite{Kubiznak2012}. Below the critical point, black holes undergo a first-order phase transition from small to large black holes, analogous to the liquid-gas phase transition observed in van der Waals fluids. The coexistence line separating the small and large black hole phases can be precisely determined by applying Maxwell's equal area law~\cite{Spallucci2013}. This line not only characterizes the phase transition dynamics between small and large black holes but also delineates the distinct single-phase regions corresponding to the small black hole phase and the large black hole phase. Similar to the gas-liquid phase transition, certain thermodynamic state functions (e.g. free energy) are continuous, but not analytic when the system is cooled along a path that crosses the coexistence line. The thermodynamic nature of AdS black holes in the coexistence region has been studied in~\cite{Wei2023}. However, in real experimental scenarios, such discontinuity/non-analyticity may not occur precisely at the coexistence line. This is because the system can persist in a supercooled metastable phase until it reaches a limit of stability, known as a spinodal line. This behavior is illustrated by path $a$ in the left panel of   Fig.~\ref{fig1}.

To describe the continuous evolution of thermodynamic state functions between different single phases (either theoretically or experimentally), the Widom line is a very useful concept~\cite{Widom1972}. This line, which extends above the critical point of the gas-liquid phase transition, describes behaviour in the supercritical regime. When a system is cooled isobarically along a path above the critical point, the thermodynamic state functions transition smoothly from the characteristic values of a high-temperature phase (e.g., a gas-like phase or a large black hole-like phase) to those of a low-temperature phase (e.g., a liquid-like phase or a small black hole-like phase). The Widom line delineates where this smooth transition takes place. This behavior is illustrated by path $b$ in the right panel of the Fig.~\ref{fig1}.

It is well established that above the critical point, the system enters the supercritical region in the phase diagram, where the distinction between the gas and liquid phases becomes ambiguous or entirely blurred. The Widom line is commonly interpreted as an extension of the coexistence curve into the one-phase region (the two crossover lines intersect at a critical point), serving as a boundary that delineates the supercritical liquid-like and supercritical gas-like phases (Visually, it can also be seen as the echo of the first-order gas-liquid phase transition in the supercritical region). Originally, the Widom line was introduced as the locus of maximum correlation length. However, for practical experimental purposes, it is often approximated as the set of states where thermodynamic response functions---such as  heat capacity at constant pressure and isothermal compressibility---exhibit extrema~\cite{Xu2005,Simeoni2010}. Notably, there is no rigorous method to directly determine the Widom line as the locus of maximum correlation length solely from these response functions. From the perspective of thermodynamic geometry, the Widom line is also closely linked to thermodynamic curvature, which is proportional to the correlation volume, for the  van der Waals model~\cite{May2012,Ruppeiner2012} and Reissner-N\"{o}rdstrom AdS fluid~\cite{Bairagya2020}.

How can we conceptualize the Widom line within the context of black hole thermodynamic systems? Unlike ordinary fluid systems, where the Widom line is typically defined using thermodynamic response functions---particularly the maximum of the constant pressure heat capacity~\cite{Xu2005,Simeoni2010}---the situation for black holes is more nuanced. In black hole thermodynamics, the charged AdS black hole serves as the most representative system. Upon entering the supercritical region,  heat capacity at constant pressure does exhibit extremal behavior; however, these extrema are strictly local, with no global maximum present. This behaviour is in stark contrast with that of van der Waals fluid systems, rendering the use of thermodynamic response function maxima unsuitable for defining the Widom line in black hole systems. 
In this study, we notice the fact that Lee-Yang phase transition theory provides a robust framework for analytically describing phase transitions in thermodynamic systems. Given that Lee-Yang zeros on the real axis correspond to phase transition points, we propose leveraging Lee-Yang phase transition theory to introduce and define the Widom line for black hole thermodynamic systems:
\begin{equation}\label{wline}
\text{Widom line}\quad \mapsto \quad \text{Projection of complex Lee-Yang zeros on the real phase plane}.
\end{equation}

Using this deductive definition, we can procure a detailed and intricate phase diagram of a black hole thermodynamic system, from which we can derive its pertinent behaviour  within the supercritical region. Analogous to the distinction between the small black hole phase (liquid phase) and the large black hole phase (gas phase) in the critical region, the Widom line allows us to introduce and differentiate between a small black hole-like phase and a large black hole-like phase in the supercritical region. This differentiation is facilitated by the Widom line, which enables a thermodynamic crossover between single-phase states without traversing the spinodal line that emerges in the critical region. Combining our definition of the Widom line in this paper and the experimental measurements of  Lee-Yang zeros in Refs.~\cite{Wei2012,Peng2015}, it provides us with the possibility of simulating (and ultimately experimentally detecting)
the critical point of thermodynamic phase transitions in AdS black holes.

We now employ concrete examples to systematically examine  supercritical behaviour in black hole thermodynamics, thereby establishing a refined phase diagram that comprehensively characterizes transition patterns for thermodynamic crossover lines.

\section{Charged AdS black hole}

In Schwarzschild-like coordinates, the metric of the four-dimensional charged AdS black hole is
$$\text{d}s^2=-f(r)\text{d}t^2+\frac{\text{d}r^2}{f(r)}+r^2\left(\text{d}\theta^2+\sin^2 \theta \text{d}\phi^2\right),$$ where  $\displaystyle f(r)=1-2M/r+r^2/l^2+Q^2/r^2$, with $M$ the Arnowitt-Deser-Misner(ADM) mass, $Q$ the total charge of the black hole, and $l$ the AdS radius. The black hole event horizon is given by the larger root of  $f(r_h)=0$, yielding $\displaystyle M=r_h/2+4\pi P r_h^3/3+Q^2/(2r_h)$, where $P$ is thermodynamic pressure via $P=3/(8\pi l^2)$ in the framework of black hole chemistry \cite{Kubiznak2017} and
the extended phase space, with $M$  the enthalpy. Using the Euclidean trick, we can identify the black hole temperature $\displaystyle T=1/(4\pi r_h)+2Pr_h-Q^2/(4\pi r_h^3)$, and the corresponding entropy is $S=\pi r_h^2$. Hence the Gibbs free energy is
$\displaystyle G=r_h/4-2\pi P r_h^3/3+3Q^2/(4r_h)$.
In addition, some thermodynamic response functions, such as constant pressure heat capacity and constant volume heat capacity, are
$\displaystyle C_P=2S(8PS^2+S-\pi Q^2)/(8PS^2-S+3\pi Q^2)$ and $C_V=0$~\cite{Kubiznak2012}.

In the extended phase space of the black hole thermodynamics, the charged AdS black hole undergoes a small-large black hole phase transition, where the critical point is
\begin{equation}\label{crp}
T_c=\frac{\sqrt{6}}{18\pi Q}, \quad P_c=\frac{1}{96\pi Q^2}, \quad r_c=\sqrt{6}Q, \quad G_c=\frac{\sqrt{6}Q}{3}, \quad C_{P(c)}\rightarrow\infty.
\end{equation}
For   convenience we  introduce the dimensionless reduced parameters $p=P/P_c$, $t=T/T_c$, $z=r_h/r_c$, $g=G/G_c$, and $c_p=C_P\cdot P_c$,
indicating that we can obtain some dimensionless reduced thermodynamic quantities
\begin{eqnarray}\label{req}
  t &=& \frac{3p}{8}z+\frac{3}{4z}-\frac{1}{8z^3}, \label{rt}\\
  g &=& \frac34\left(z-\frac{p}{6}z^3+\frac{1}{2z}\right), \label{rg}\\
  c_p &=& \frac{z^2(3pz^4+6z^2-1)}{24(pz^4-2z^2+1)},\label{rcp}
\end{eqnarray}
and the small/large coexistence line  below the critical point is~\cite{Spallucci2013}
\begin{equation}\label{coel}
t=\sqrt{\frac{p(3-\sqrt{p})}{2}},
\end{equation}
which starts from the origin and ends at the critical point~(\ref{crp}).

\begin{figure}[htbp]
	\centering
		\includegraphics[width=57 mm]{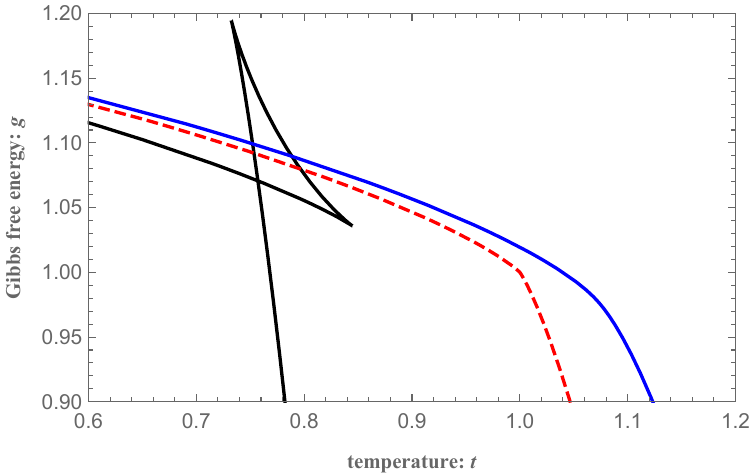}
        \includegraphics[width=55 mm]{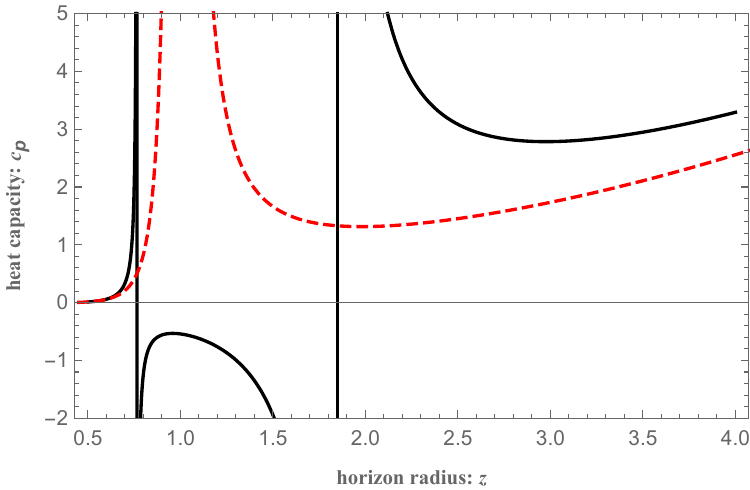}
        \includegraphics[width=55 mm]{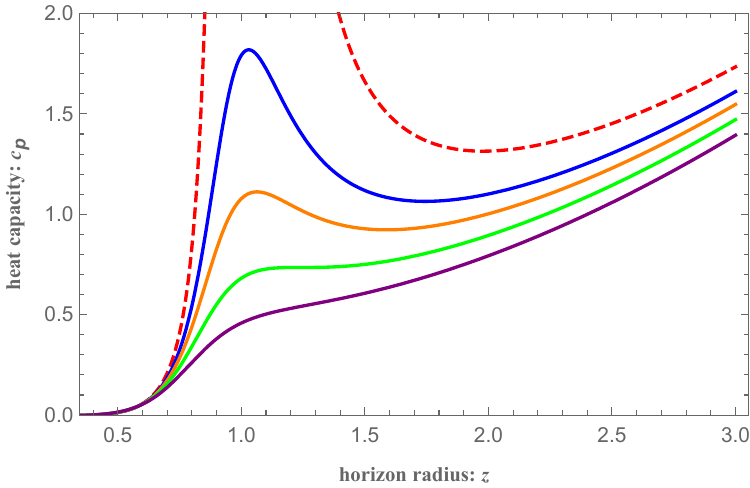}
	\caption{The behaviors of dimensionless reduced Gibbs free energy $g$ and constant pressure heat capacity $c_p$, where black line for $p=0.50$, red dashed line for $p=1.00$ (critical point), blue line for $p=1.20$, orange line for $p=1.35$, green line for $p=1.60$, and purple line for $p=2.00$ for the charged AdS black hole.   }
	\label{fig2}	
\end{figure}

We depict the behavior of the Gibbs free energy $g$~(\ref{rg}) and constant pressure heat capacity $c_p$~(\ref{rcp}) for different values of the pressure in Fig.~\ref{fig2}. It can be seen that below the critical point, the Gibbs free energy exhibits some non-analyticity, namely at the two cusps of the swallowtail structure. These two points correspond to  divergent behavior in $c_p$. After exceeding the critical point, $g$ is continuous and monotonic.  $c_p$ is also a continuous function that has some local
extrema. As $p$ continuously increases, we see that the extreme behaviour of   $c_p$ disappears and it becomes a monotonically increasing function.

\begin{figure}[htbp]
	\centering
		\includegraphics[width=120 mm]{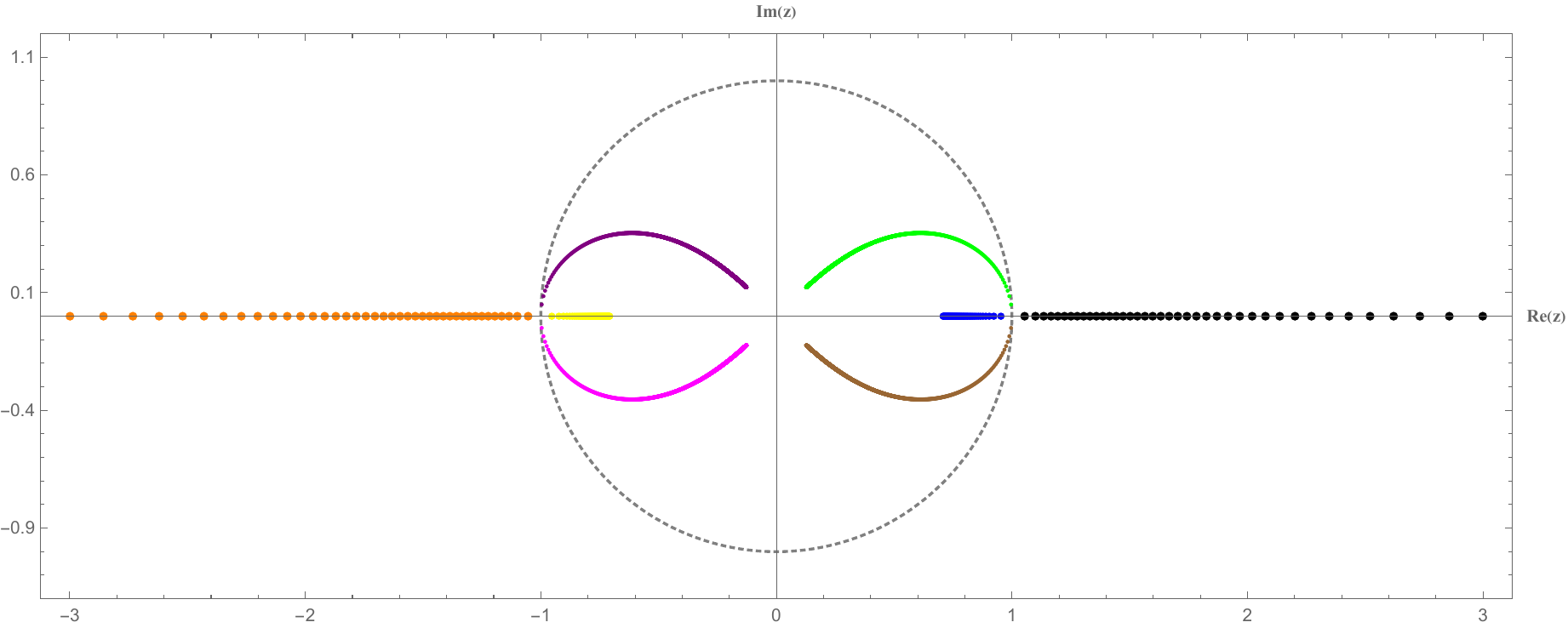}
	\caption{The singularity distribution of the Gibbs free energy in charged AdS black hole, corresponding to the Lee-Yang zeros, for different pressure values. The gray curve represents the unit circle, while distinct colors denote the four roots of Eq.~(\ref{fenbu}) in the complex domain. Specifically, roots located on the real axis correspond to the critical region ($p<1$), whereas those in the complex plane (excluding the real axis) are associated with the supercritical region ($p>1$).  There are also singularities in the second derivative of $g$ at $z=\pm 1$, on the unit circle.
	}
	\label{fig3}	
\end{figure}

We now extend our analysis to the complex domain to investigate the thermodynamic behavior of charged AdS black holes in the supercritical region. In this framework, we treat $z$ as a complex variable, which consequently makes $t$ a complex variable as well, while $p$ remains a real number~\cite{Xu2024b}. As established in prior analyses, a phase transition in a black hole thermodynamic system is associated with the singularity distribution of the Gibbs free energy. From the left  diagram in Fig.~\ref{fig2}, by examining the behavior of the Gibbs free energy as a function of temperature, it becomes evident that the non-analyticity of the free energy is manifest as two distinct cusps within the swallowtail structure. At these points, the second derivative of the free energy~(\ref{rg}) with respect to temperature exhibits a discontinuity. Hence we obtain the singularity distribution of the Gibbs free energy, i.e., the Lee-Yang zeros according to Eq.~(\ref{rule}),
\begin{equation}\label{fenbu}
pz^4-2z^2+1=0.
\end{equation}
For values of $p$ we depict the singularity distribution of the Gibbs free energy in Fig.~\ref{fig3}. We see that these singularities are divided into two categories. One type is located on the real axis, where $p<1$. Those singularities located on the positive real axis correspond to real phase transition points. The singularities of other type are distributed in the complex plane, corresponding to supercritical phenomena with $p>1$.

The distribution of Lee-Yang zeros provides crucial insight into the thermodynamic phase behaviour of the system. When no genuine phase transition occurs in the thermodynamic system, as evidenced by the Lee-Yang zeros being distributed exclusively in the complex plane (excluding the real axis), these zeros are consistently observed to reside within the unit circle. However, upon the occurrence of a phase transition, the Lee-Yang zeros extend beyond the unit circle and distribute along the positive real axis. This distribution pattern fundamentally reflects the analytic properties of the system's phase transition. Notably, based on our previous proposition, the zeros located in the complex plane (excluding the real axis), particularly those in the first quadrant, serve as critical indicators for understanding the supercritical phenomena of the system.

\begin{figure}[htbp]
	\centering
		\includegraphics[width=80 mm]{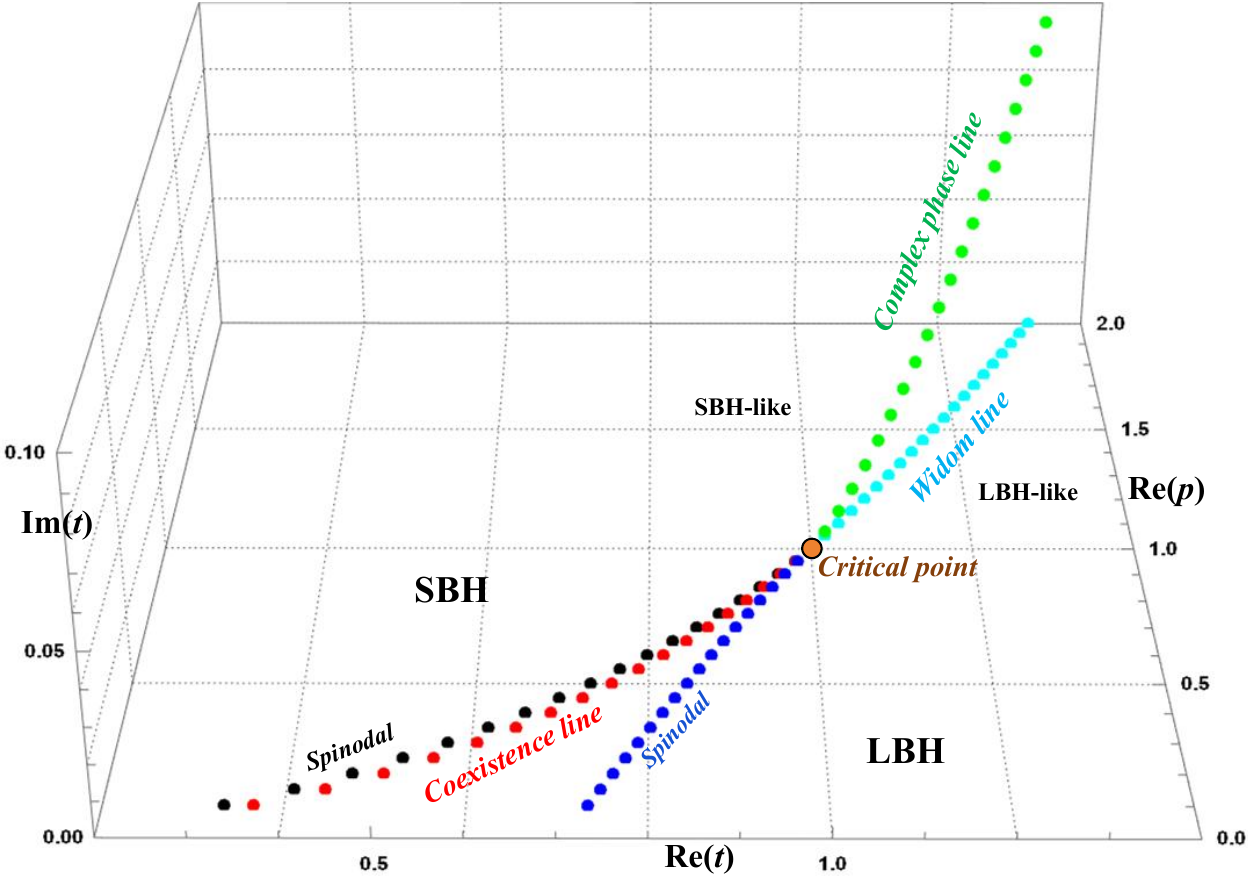}
	\caption{The complex phase diagram of the charged AdS black hole in a three-dimensional complex space (consisting of the positive real part of temperature $\text{Re} ~t$, the positive real part of pressure $\text{Re} ~p$, and the positive imaginary part of temperature $\text{Im} ~t$) and corresponding supercritical phenomena.}
	\label{fig4}	
\end{figure}

Based on Eqs.~(\ref{rt}),~(\ref{coel}) and~(\ref{fenbu}), we   obtain the phase diagram of a charged AdS black hole in the complex domain, shown in Fig.~\ref{fig4}. From this, we can infer more diverse phase transition characteristics of the black hole thermodynamic system.

\begin{itemize}
  \item The coexistence line (red line) between small black hole and large black hole is determined by Eqs.~(\ref{coel}), which originates from the zero point and terminates at the critical point. If the system is subject to a cooling process under constant pressure conditions,  a phase transition occurs  as it traverses the coexistence line, characterized by a shift from the large black hole state to the small black hole state. This transition is accompanied by distinct discontinuities in the thermodynamic state functions.
  \item The spinodal lines (black and blue lines, respectively corresponding to those in Fig.~\ref{fig3}) correspond to singularities of the Gibbs free energy located on the positive real axis in Fig.~\ref{fig3} for $p<1$. There exist two spinodal lines, and the regions between them and the coexistence line correspond to the super-cooled and the super-heated metastable phases, respectively.
   \item The complex phase line (green line, corresponding to the green line in Fig.~\ref{fig3}) is determined by the singularities of the Gibbs free energy in the first quadrant of the complex plane in Fig.~\ref{fig3}. It is a new phase diagram line residing in a three-dimensional complex space, which reflects the analytic properties of phase transitions and encapsulates significant information about the system in the supercritical region.
   \item The projection of the complex phase line onto the real phase plane is known as the Widom line (cyan line, based on Eq.~(\ref{wline})), and falls within the supercritical region which emanates from the critical point. This crossover line divides the phase diagram above the critical point into regions with distinct physical characteristics, namely, the small black hole-like phase and the large black hole-like phase. These regions can be connected without encountering any thermodynamic singularities. The Widom line is generally finite in extent. According to its rough definition in supercritical fluids---as the locus of maxima in isobaric heat capacity---the extremal behavior of these response functions gradually disappears with increasing temperature or pressure, indicating termination of the Widom line. This conclusion similarly applies to our definition based on Lee-Yang zeros, where the Widom line also terminates at a specific point. However, our primary research interest focuses on the properties near the critical point, particularly the finite segment of the Widom line immediately above the critical point. The central question involves how to properly define phase boundaries within this limited supercritical region beyond the critical point.
   \item The other lines in Fig.~\ref{fig3},  located in the second to fourth quadrants of the complex plane,   correspond to cases where the real or imaginary part of $z$ is negative, in which case the  real or imaginary part of the corresponding complex temperature is respectively negative. We therefore do not consider these physically meaningless situations and limit our analysis to the case presented in the first quadrant of Fig.~\ref{fig3}.
\end{itemize}

In the supercritical region of a physical system, the distinction between two phases (say gas and liquid) becomes blurred, and the relevant real thermodynamic potentials and various response functions are continuously changing. In this region the Widom line is the boundary between  supercritical liquid-like and supercritical gas-like phases, despite the absence of a phase transition between them, and  the monotonic behavior of some thermodynamic response functions is different on both sides. For this  reason the Widom line is usually defined using the extremum of a thermodynamic response function. However for black holes, this is not adequate. For the charged AdS black hole we consider, according to the right diagram in Fig.~\ref{fig2}, we see that the slope of the constant pressure heat capacity is divided into three parts -- ``positive-negative-positive'' --  which leads to an ambiguity in using the extreme value of the response function to define the Widom line.

\begin{figure}[htbp]
	\centering
		\includegraphics[width=65 mm]{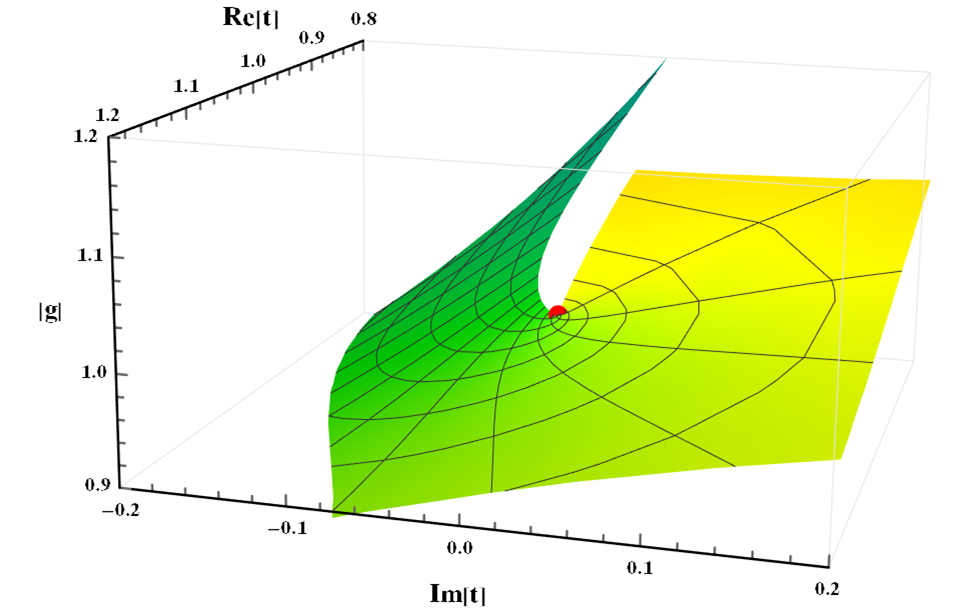}
        \includegraphics[width=65 mm]{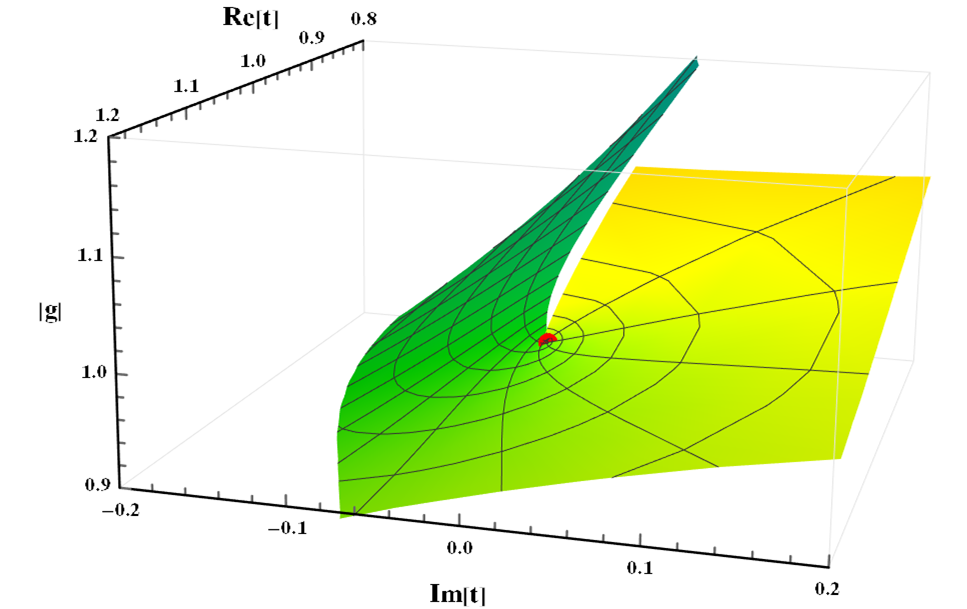}
        \includegraphics[width=65 mm]{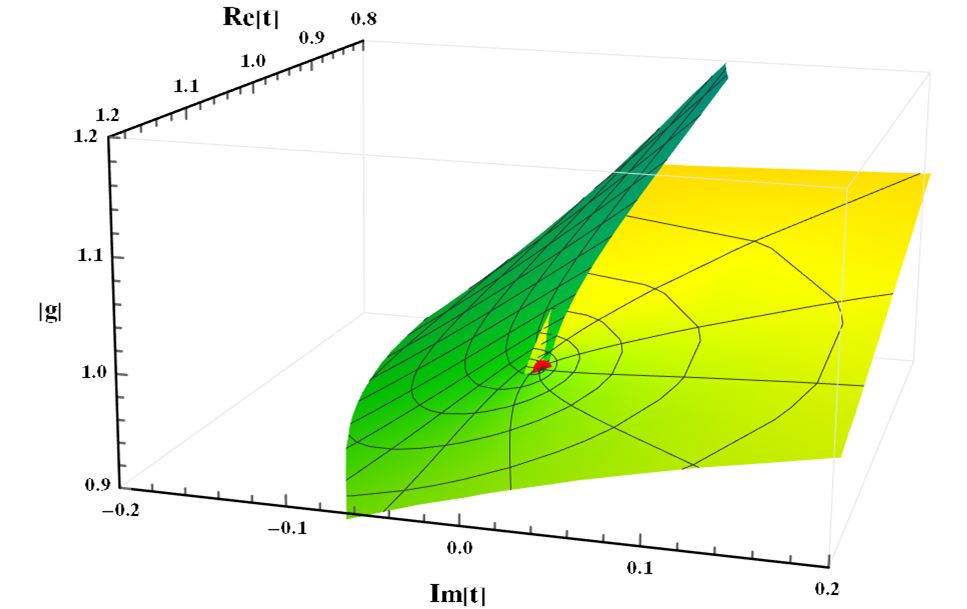}
        \includegraphics[width=65 mm]{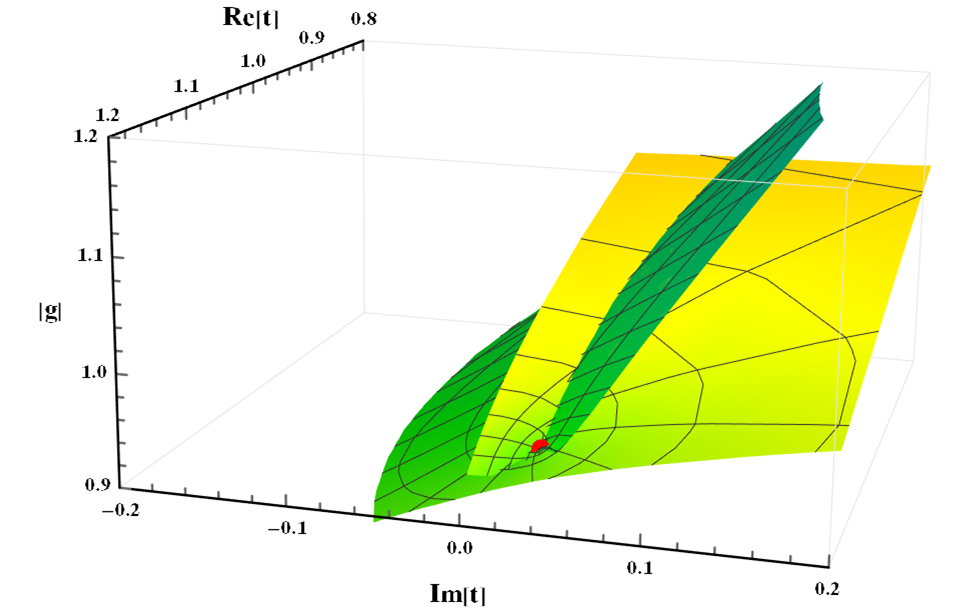}
	\caption{The behaviors of moduli of the complex Gibbs free energy $g$ with respect to complex temperature $t$ for different pressure $p=1.0$ (upper-left), $p=1.1$ (upper-right), $p=1.2$ (lower-left), $p=1.5$ (lower-right) for the charged AdS black hole, where the red dot represents the Widom point (line) at this pressure, which is the boundary point (line) between the two phases in the supercritical region.  }
	\label{fig5}	
\end{figure}

For the scheme we propose, we use the Lee-Yang transition theorem to derive the Widom line in the supercritical region by projecting the complex Lee-Yang zeroes onto the real plane. To extend the analysis to a complex field, we consider complex temperature, which leads to complex Gibbs free energy. Clearly measurable physical quantities can be obtained from  the moduli of these complex functions. In   Fig.~\ref{fig5}, we plot the relationship between the modulus of the complex Gibbs free energy with respect to the complex temperature at different pressures. It can be clearly seen that   the modulus of the complex free energy is a smooth surface. As the pressure exceeds the critical point and continues to increase, the two edges of this surface gradually approach and intersect, a process that revolves around the Widom point (line). This decomposes the originally smooth surface into two foliations. On either side of the Widom point (line), the moduli of the free energy are distinctly discontinuous, indicating different physical properties. Consequently, two distinct phases are delineated by means of the Widom line within the supercritical region.

\section{Summary}

The precise definition of the boundary line in the supercritical region (analogous to the coexistence line in subcritical systems) remains an open question, with each existing definition having its own theoretical justification. However, the experimentally observed phenomena cannot be ignored: distinct fluid phases (gas-like and liquid-like) with different physical properties have been consistently identified within supercritical fluids \cite{Simeoni2010}. This observation has been verified through both experimental measurements and molecular dynamics simulations. The current lack of comprehensive theoretical explanations for this phenomenon leaves significant room for further investigation.

Notwithstanding these issues,  we have employed  Lee-Yang phase transition theory to analyze supercritical phenomena in black hole thermodynamic systems.
We have shown how to  establish a crossover line,  the Widom line, within the supercritical region of a  black hole thermodynamic system,   partitioning it into two distinct phases with disparate physical characteristics, designated as the small black hole-like phase and the large black hole-like phase. When the system is cooled through an isobaric process within the supercritical region, crossing the Widom line signifies a transition from one supercritical phase to another, during which the thermodynamic state functions of the system vary continuously, whereas the modulus of the complex free energy, which located on two foliations, is distinctly discontinuous.

Adhering to the core idea of the Lee-Yang phase transition theorem, where it is well established that real Lee-Yang zeros correspond to phase transitions, we propose that complex zeros - particularly those projected onto the positive real plane - can equally serve as valid indicators of phase separation. Notably, the real projections of these complex zeros naturally reside in the supercritical region of the phase diagram and maintain smooth continuity with the coexistence line at the critical point. This characteristic makes them suitable for defining the Widom line in the supercritical regime with the fundamental correspondence through Eqs.~(\ref{rule}) and~(\ref{wline}). To our knowledge, this study represents the first application of Lee-Yang phase transition theory to black hole thermodynamics, introducing a complex-analytic approach to characterize phase transitions in such systems. By interpreting black hole phase transitions through the lens of Lee-Yang zeros, we provide new theoretical insights that advance our understanding of black hole physics, at least at the fundamental theoretical level.

\section*{Acknowledgments}
This research was supported in part by the Natural Sciences and Engineering Research Council of Canada and by the National Natural Science Foundation of China (Grant No. 12575064, No. 12247103, and No. 12105222) and by Natural Science Basic Research Plan in Shaanxi Province of China (Grant No. 2025JC-YBQN-029) and supported by the project of Tang Scholar in Northwest University. The authors would like to thank the anonymous referee for  helpful comments.
\section*{Note added}Upon completion of our work we became aware of ref.~\cite{Zhao2025} in which the Widom Line was defined for the charged AdS black hole inspired by the higher-order thermodynamic derivatives (variance/skewness/kurtosis) with dynamical verification through distinct quasi-normal modes. Both teams independently arrived at identical conclusions through distinct methodologies. The boundary line derived from the charged AdS black hole’s supercritical region provides compelling evidence for a supercritical phase transition.

\end{document}